# Distributing entanglement and single photons through an intra-city, free-space quantum channel

K.J. Resch[1,*], M. Lindenthal[1,*], B. Blauensteiner[1], H.R. Böhm[1], A. Fedrizzi[1], C. Kurtsiefer[2,3], A. Poppe[1], T. Schmitt-Manderbach[2], M. Taraba[1], R. Ursin[1], P. Walther[1], H. Weier[2], H. Weinfurter[2], and A. Zeilinger[1,4]

[1]*Institut für Experimentalphysik, Universität Wien, Boltzmanngasse 5, A-1090 Wien, Austria*
[2]*Sektion Physik, Ludwig Maximilians Universität, D-80797, München, Germany*
[3]*Quantum Information Technology Lab, National University of Singapore, 2 Science Drive 3, 117542 Singapore*
[4]*IQOQI, Institut für Quantenoptik und Quanteninformation, Österreichische Akademie der Wissenschaften, Austria*
*These authors contributed equally to this work
resch@quantum.at   michael.lindenthal@quantum.at

**Abstract:** We have distributed entangled photons directly through the atmosphere to a receiver station 7.8 km away over the city of Vienna, Austria at night. Detection of one photon from our entangled pairs constitutes a triggered single photon source from the sender. With no direct time-stable connection, the two stations found coincidence counts in the detection events by calculating the cross-correlation of locally-recorded time stamps shared over a public internet channel. For this experiment, our quantum channel was maintained for a total of 40 minutes during which time a coincidence lock found approximately 60000 coincident detection events. The polarization correlations in those events yielded a Bell parameter, S=2.27±0.019, which violates the CHSH-Bell inequality by 14 standard deviations. This result is promising for entanglement-based free-space quantum communication in high-density urban areas. It is also encouraging for optical quantum communication between ground stations and satellites since the length of our free-space link exceeds the atmospheric equivalent.

## References and links

## 1. Introduction

Most quantum communication protocols require communicating parties to transmit either single qubits or entanglement between distant locations [1]. Of all of the systems experimentally investigated for quantum information, only photons, traveling at the speed of light and being relatively immune to decoherence, seem suited to this task. Optical fibre and free-space links have proven potential for low-loss distribution of photons over long distances. Entangled photons have been distributed in optical fibre for long-distance fundamental tests or quantum cryptographic applications [2-9]. Free-space links have been used in long distance quantum cryptography experiments using faint laser pulses [10-14] and recently were used to distribute entangled photons over 600m [15]. While the experimental progress in this area has been rapid, theoretical studies have shown that Earth-bound fibre and free-space quantum communication based on current technology cannot surpass on the order of 100km [16]. An exciting alternative that is within reach is a proposal to use free-space links with orbiting satellites which circumvent the problems of short lines-of-sight and high atmospheric scattering here on Earth [17-20]. We have built up a quantum communication experiment where the source is placed inside a 19th-century observatory, Kuffner Sternwarte (Alice), and a receiving station is 7.8km away on the 46th floor of a modern skyscraper, Millennium Tower (Bob). The sending telescope is located 15m above the ground on the roof of the observatory and the receiver is located 150m above ground. The latter height is representative of the height of the beam over most of the propagation path as the observatory is located on a hill. Cities are challenging environments for free-space optical communication since pollution and atmospheric turbulence lead to high scattering and beam fluctuations, both of which are detrimental to link efficiency and stability. Our free-space link was intentionally chosen to pass over the heart of Vienna, Austria to prove the potential for quantum communication in the environment where it is most likely to find application. Our free-space link distance corresponds to just over one airmass, the characteristic thickness of the Earth's atmosphere, of 7.3km [21]. It is also much longer than 4.5 km which is the horizontal distance where scattering loss is the same as through the whole atmosphere vertically [22]. This experiment, therefore, serves to test the feasibility of sending entangled photons through free-space optical links under atmospheric conditions as close as possible to those that will be encountered in future space experiments.

In the present experiment, we used a source emitting polarization-entangled photon pairs from parametric down-conversion. In addition to their polarization entanglement, down-converted photon pairs are emitted with very tight temporal correlations. It was recognized early on that these time correlations allowed low-noise communication based on coincidence events in a narrow time window even when the background was larger than the signal intensity [23-25]. Furthermore, the time correlations allow for the production of heralded single-photons if one of the two photons is detected locally at the source and information about this detection event is broadcast to the receiver [26]. In most experiments using entangled photons, coincident pairs are identified by comparing detection events over time-stable classical channels. This channel is most commonly in the form of a direct cable connection, although in some long-distance experiments involving entangled photons or faint laser pulses, additional bright optical pulses have been used to carry detection and timing information [9,11,25]. Here, we use no such direct connection and rely only on publicly-shared internet as our classical communication channel. Using an experimental technique applied to closing the locality loophole in a Bell test [6], the times at which each detection event occurs are recorded and shared over the internet. Coincident events are determined using their cross-correlation. In essence, the narrow temporal correlations in photon number from the down-conversion process provide timing stability. In order to demonstrate that the photon pairs shared by Alice and Bob are truly entangled, we measured their polarization correlations for a test of the CHSH-variant of Bell's inequalities [27,28]. The coincidence counts extracted from our measured time-tags were used to demonstrate a convincing violation of this inequality by 14 standard deviations. This result confirms the high quality of our long-distance free-space quantum channel and is promising for future quantum communication experiments in city environments and between ground and space.

## 2. Experimental

The basic layout of this real world experiment is shown in Fig. 1. Alice is located at the Kuffner Sternwarte (KSW) where a portable source of entangled photons is placed inside on an optical breadboard. The source is pumped by a 15-mW, 405-nm violet diode laser which produced photon pairs at 810nm via type-II spontaneous parametric down-conversion (SPDC) in a $\beta$-barium borate crystal (BBO) [29]. The coincident photons have a bandwidth of 5.8nm FWHM. These photons pass through additional compensation BBO crystals which offset transverse and longitudinal walk-off effects. The tilt angle of these compensation crystals are used to change the relative phase between horizontal (H) and vertical (V) polarizations for aligning the source to produce photon pairs in the singlet state,

$$|\psi^-\rangle = \frac{1}{\sqrt{2}}(|H\rangle_A|V\rangle_B - |V\rangle_A|H\rangle_B), \qquad (2)$$

where the subscripts label the spatial modes. Each photon is coupled into a single-mode optical fibre, a section of which contains a polarization controller used to counteract fibre-induced polarization rotations. Each detection module used in this experiment has four output channels. Incident photons first pass a 50/50 beam-splitter where light in the reflected path is measured in the H/V (i.e., 0°/90°) basis and transmitted light is measured in the +/- (i.e., 45°/135°) basis. In both bases, measurements are made by polarizing beam-splitters (PBS) and photons are counted using avalanche photo-diodes. The photon in mode A is detected directly on Alice's breadboard using such a multi-channel detector. After the polarization optics, the light is detected using avalanche photo-diodes. The photodiodes used for the experiment at Alice are passively quenched and not fibre coupled, those used at Bob for the experiment are actively quenched and multi-mode fibre coupled. The dark counting rate for each passively-quenched and actively-quenched detector channel is 1200 s$^{-1}$ and 800 s$^{-1}$, respectively. Measuring both photons locally with Alice's four channel detection module and one actively quenched fibre-coupled detector, our source produces entangled photon pairs with a measured coincidence rate of 9000 s$^{-1}$ without narrowband interference filters. The detection efficiency of the source, which also includes the coupling efficiency into the single-mode optical fibres is approximately 15% for the actively-quenched detectors and 9% for the passively-quenched detectors. Also measured locally, the polarization correlations have 96% visibility in the H/V basis and 90% in the +/- basis demonstrating the high-quality of the entanglement.

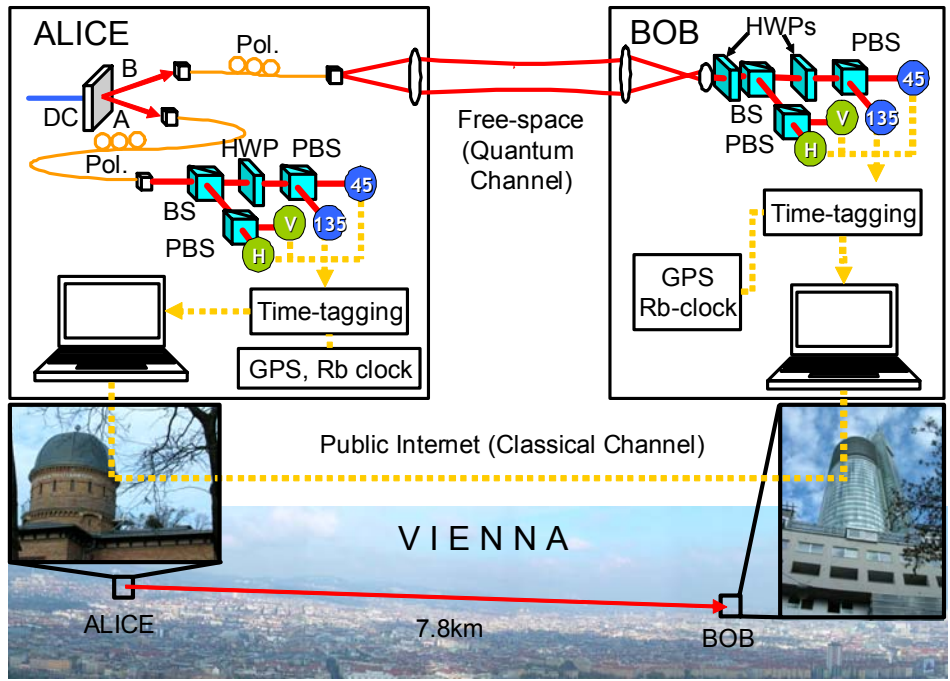

Fig. 1. The layout for free-space entanglement distribution in Vienna. Alice has the single-mode fibre coupled polarization-entangled photon source (DC) and sending telescope and is located in the 19th century observatory, Kuffner Sternwarte. Bob has a receiver telescope and is located on the 46th floor of the Millennium Tower skyscraper 7.8km away. Alice measures the photons in mode A from each entangled pair using a four-channel detector, made with a 50/50 beam-splitter (BS), a half-wave plate (HWP), and polarizing beam-splitters (PBS), which measures the photon polarization in either the H/V or +/- basis. She sends the other photon in mode B, after polarization compensation (Pol.), via her telescope and free-space link, to Bob. Bob's receiver telescope is equipped with a similar four-channel detector and can measure the polarizations in the same bases as Alice or, by rotating an extra HWP, measure another pair of complementary linear polarization bases. Alice and Bob are both equipped with time-tagging cards which record the times at which each detection event occurs. Rubidium atomic clocks provide good relative time stability. Both stations also embed a 1pps signal from the global positioning system (GPS) into their time-tag data stream to give a well-defined zero time offset. During accumulation, Bob transmits his time tags in blocks over a public internet channel to Alice. She finds the coincident photon pairs in real time by maximizing the cross-correlation of these time tags. Which of the four detector channels fired is also part of each time tag and allows Alice and Bob to determine the polarization correlations between their coincident pairs. Alice uses her polarization compensators to establish singlet-like anti-correlations between her measurements and Bob's.

In the experiment, the photon in mode B is not detected locally, rather it is sent via a 10-m single-mode fibre from the source table inside the observatory to a platform constructed on the roof. From here, the light emerges from the fibre, expands, and is collimated by a 15-cm-diameter achromatic lens (f = 40cm) into an approximately 8-cm diameter beam. The tip and tilt of the telescopes are controlled from inside to a precision better than 10μrad using stepper motors.

After propagating the 7.8-km line-of-sight distance from Kuffner Sternwarte to the Millennium Tower, the photons arrive at Bob's station. Before connecting the down-conversion source, a 2-mW, fibre-coupled diode laser is used to align the telescopes. On the night of the experiment, this produced a spot at the receiver of about 25cm in diameter that typically wandered an additional 25cm due to atmospheric turbulence. Atmospheric conditions can dramatically affect the beam characteristics; beam spot sizes and wander have been observed as small as 15cm and as large as 1m. The receiver lens, identical to that of the sender, focuses the light to a small spot. In its focal plane, a 150-μm diameter pinhole acts as a spatial filter with a 200-μrad half-angle acceptance, corresponding to a 3-m diameter spot at the source, for background suppression. After the pinhole, the light is collimated by a second lens and passes through the second detector module with four output channels similar to that used at Alice. However, after the PBSs the light is coupled into multi-mode optical fibres in order to increase the collection efficiency. Those multimode fibres can be connected either to a power meter during the alignment procedure with visible laser beams, or connected to multimode-fibre-coupled single-photon counting detectors. A half-wave plate can be inserted before the 50/50 beam-splitter so that the light could be rotated before entering the polarization analyzer. By rotating the wave-plate from 0° to 11.25° we rotate the polarization of the light by 22.5° and simultaneously perform all of the polarization measurements required for testing the CHSH-Bell inequality.

In this experiment, we determine coincident events using time-stamping techniques instead of a time-stable channel. Time-stamping cards record not only that an event has occurred, but at which time that event occurred with a precision of 125ps. Each time-stamp is recorded as a 64-bit binary number. The time-stamping cards are stabilized by Rubidium oscillators with a relative drift of about 1ns every 20s. In addition to recording the time at which an event occurs, the time-stamping cards also encode which of the 4 channels fired. To give both time-stamping cards a well-defined absolute time reference (i.e., a common reference for zero time), we embed a simulated 8-fold coincidence count into the data stream using the pulse-per-second (pps) signal from the global positioning system (GPS). With our experimentally-observed counting rates, a real 8-fold coincidence count occurs with a negligible rate (about $10^{-28}$ s$^{-1}$). This 1-pps signal provides an absolute time reference with an uncertainty of 50ns. To start the measurement at approximately the same time at both Alice and Bob to within $10^{-2}$ s, each computer at the two locations use the network timing protocol (NTP) to synchronize their internal clocks to a server. The ultimate time stability comes from the time correlations and coincident emission of the down-converted photons themselves.

Time stamps at each station are locally collected. For real-time coincidence calculation, Bob sends his time-stamp data in 64-kb blocks over a standard internet connection to Alice.

Due to the loss in the free-space link, Bob measures fewer time-stamps and therefore it is much more economical to send the time stamps in this direction. The bandwidth between Alice and Bob is about 1Mb/s corresponding to a maximum rate of about 15000 time stamps/s. This bandwidth is sufficient to exchange the 6500-8500 counts/s measured. Initially, the software locates the GPS 1pps embedded in the data stream that provides a zero time to begin the coincidence search. From there, the cross-correlation of the two lists of time stamps is calculated. A given shift between the data files is registered as the correct shift for coincidence counts once the cross-correlation at that shift reaches a controllable multiple of the accidental coincidence rate. Once found, the coincidence rate and all 16 correlations between the different detectors are displayed in real time. The unwanted polarization rotations incurred in the single-mode fibres at Alice are compensated using these online correlations to yield strong singlet-like correlations between source and receiver.

## 3. Results

In the experiment, the receiver collected approximately 2% of the infra-red diode laser light sent from the source table. Switching to entangled photons from the down-conversion source, we measured a singles rate of 7200 $s^{-1}$ of which 5850 $s^{-1}$ were background. Note that the dark counting rate of the four actively-quenched diode channels was 3200 $s^{-1}$. The background is the sum over all four detector channels and is achievable under nighttime conditions using only spatial filtering in the telescope with no additional spectral filtering. (When the spotlight at the observatory was on, the background level increased to over 50 000 $s^{-1}$ even though this light was insufficient to make the observatory visible to the eye from Millennium Tower). Our singles rate measured at the source using the same types of single-photon counting modules as those in the receiver was 100 000 $s^{-1}$. Thus our collection efficiency for the light from the source was, on average, 1.4%, with large fluctuations originating from atmospheric turbulence. Note that this efficiency includes the losses from the sender lens, receiver lens, polarization optics and fibres, but not the efficiency of the single-photon counting detectors themselves. This total link efficiency is in reasonable agreement with the efficiency obtained with the laser diode.

Table 1. Experimentally-measured coincidence rates.

| | | Millennium Tower (Bob) | | | |
|---|---|---|---|---|---|
| | | 22.5° | 112.5° | 67.5° | 157.5° |
| Kuffner Sternwarte (Alice) | 0° | 1469 | 5763 | 6500 | 1067 |
| | 90° | 4015 | 1305 | 1483 | 2959 |
| | 45° | 2171 | 9103 | 2633 | 6357 |
| | 135° | 1701 | 1701 | 6889 | 1090 |

Each receiver had four output channels which corresponded to four different polarization measurements in two different bases. The polarization of the photons measured locally at Kuffner Sternwarte were either measured in the 0°/90° (horizontal/vertical) basis or the 45°/135° basis. The polarizations of the photons measured at Millennium Tower were measured in the 22.5°/112.5° or 67.5°/157.5° bases. The coincidence lock was maintained for a total of 715s during approximately 40 minutes of data taking. The coincidence window was selected to be 7ns by our cross-correlation program. We stress that these counts were directly measured in the experiment with no form of background or noise subtraction. These 16 coincidence rates, which were all measured together in a static experimental setup, can be used to give the necessary polarization correlations for a test of the CHSH-Bell inequality.

Bell's inequalities are bounds on the strength of correlations of dichotomic measurement outcomes placed by the assumptions of locality and realism [27]. The CHSH-Bell inequality [28] states that under the assumptions of local realism, the Bell parameter $S \leq 2$. The Bell parameter can be expressed as

$$S = |E(\varphi_A, \varphi_B) - E(\varphi_A, \widetilde{\varphi}_B) + E(\widetilde{\varphi}_A, \varphi_B) + E(\widetilde{\varphi}_A, \widetilde{\varphi}_B)|, \qquad (2)$$

in which $E(\varphi_A, \varphi_B)$ is the polarization correlation for analyzer settings $\varphi_A$ and $\varphi_B$ at Alice and Bob respectively. Quantum mechanics does not respect this inequality and is, thus, not a local realistic theory. Our target state, $|\psi^-\rangle$, can reach the quantum mechanical limit $S = 2\sqrt{2} \approx 2.83$ for the choice of angles $\{\varphi_A, \widetilde{\varphi}_A, \varphi_B, \widetilde{\varphi}_B\} = \{0°, 45°, 22.5°, 67.5°\}$.

After aligning the polarization compensators for maximum singlet anti-correlations in the H/V and +/- bases, we rotated the incoming polarization by 22.5° with the half-wave plate. Our experimentally measured coincidence rates from this configuration are shown in Table 1. These coincidence counts were directly extracted from the timestamp data with no form of background subtraction. They were obtained over the course of 40 minutes during which our software was able to maintain a lock on the coincidence counts for 715 seconds (~12 minutes). During this time, 59878 coincidences were recorded using a coincidence window of 7ns. Thus our average coincidence rate was 84 s$^{-1}$ during those times where we had a coincidence lock and 25 s$^{-1}$ over the full duration of the experiment. Assuming Poissonian errors on these coincidence counts, we extract the polarization correlations as shown in Table 2. The Bell parameter, S, is calculated according to Eq. (2) to be $S_{EXP} = 2.27 \pm 0.019$ - 14σ larger than the upper limit $S = 2$ imposed by local realism. This convincing violation of the CHSH-Bell inequality confirms the successful distribution of entanglement between the two city buildings 7.8 km apart through our free-space quantum channel.

Table 2. The polarization correlations between the different bases.

| $(\phi_A, \phi_B)$ | $E(\phi_A, \phi_B)$ |
|---|---|
| $(0°, 22.5°)$ | $-0.558 \pm 0.011$ |
| $(0°, 67.5°)$ | $+0.575 \pm 0.009$ |
| $(45°, 22.5°)$ | $-0.578 \pm 0.009$ |
| $(45°, 67.5°)$ | $-0.561 \pm 0.008$ |

The 16 coincidence measurements shown in Table 1 combine to yield the four polarization correlations shown. The uncertainties for each correlation have been calculated using Poissonian errors on the coincidence counts. Using these correlations yields a Bell parameter, $S_{EXP}$ = 2.27±0.019, which strongly violates the CHSH Bell inequality conclusively demonstrating the shared entanglement between Kuffner Sternwarte and Millenium City.

The measured Bell parameter allows us to quantify our quantum channel for future applications in quantum cryptography. The anticipated quantum bit error rate (QBER) can be estimated from the expression,

$$QBER = \frac{1}{2}\left(1 - \frac{S_{EXP}}{S_{QM}}\right), \qquad (3)$$

where $S_{QM}$ is the Bell parameter predicted from quantum mechanics $2\sqrt{2}$ [15]. In our case, QBER ≃ 9.9% which is low enough for secure quantum key distribution [30]. However, the secure bit rate after bases reconciliation, error correction, and privacy amplification would be at least a factor of 10 lower than the measured coincidence rate [31]. The inferred QBER corresponds to a drop in average visibility to from 93% at the source to 80% at the receiver. In the full experiment, we can attribute 3.5% QBER to the source characteristic, 1.5% QBER to accidental coincidence counts, and the remainder to polarization misalignments. With no spectral filtering, the present experiment is suited for nighttime applications; future work will investigate the necessary adaptations for low-background daylight operation.

The QBER estimation procedure assumes white noise which we most certainly did not have in the experiment. Even before the final polarization rotation, the state had observable correlations in the cross bases. It is a difficult task to compensate the polarization rotation incurred in the optical fibres at KSW under real experimental conditions which lead to low and unstable coincidence signals. It is more likely to attribute some of the loss in quality to a partially-rotated state. Thus, it is quite possible that our final state was more entangled than implied by our measured $S$ value.

### 4. Conclusions

We have distributed polarization-entangled photons using a free-space optical link over a distance of 7.8km through the heart of Vienna. Photons are identified and analyzed without a time-stable channel; rather, the time at which single-photon detection events occur are locally recorded with time tags which were subsequently shared over a public internet channel. Coincidences were found using the cross-correlation of those time tags. As such, the sending and receiver stations were completely independent. The polarization correlations contained in our measured time tags were sufficient to convincingly violate a CHSH-Bell inequality and demonstrate entanglement between our two city buildings.

We have extended the distance over which entangled photons have been distributed through free-space optical links by an order of magnitude. In doing so, we have surpassed the horizontal distance where the optical properties of the air are comparable to passing vertically through the entire atmosphere. Our results show that high-fidelity transfer of entangled photons is possible under these real-world conditions and are promising for future quantum communication using satellites.

### Acknowledgements

The authors thank W. Wondrak and L. Stohwasser for their technical assistance. The authors are grateful to P. Habison, M. Hanny, H. Neu, W. Reithofer, A. Schmidt, R. Seller, H. Steiner, G. Stumpf for their support. This work was funded by the Austrian Science Foundation (FWF) project number SFB 015 P20, the European Commission, contract no. IST-2001-38864, RAMBOQ, the DOC program of the Austrian Academy of Sciences, the ASAP programme of the Austrian Space Agency, the Federal Ministry of Transport, Innovation and Technology, project QUANTUM, NSERC, Kuffner Sternwarte, Stumpf AG (Millennium Tower), Wienerberger AG (Twin Towers), Inode GmbH, Energie AG Oberösterreich and ARC Seibersdorf research GmbH.